\documentclass[prl,twocolumn,amsmath,superscriptaddress]{revtex4}

\usepackage{graphicx}
\usepackage{dcolumn}
\usepackage{bm}

\usepackage{epstopdf}

\begin{document}
\preprint{APS/123-QED}

\title{Creating pseudo Kondo-resonances by field-induced diffusion of atomic
hydrogen}

\author{Werner A. Hofer}
\affiliation{Surface Science Research Centre, the University of
Liverpool, Liverpoool L69 3BX, UK}
\author{Gilberto Teobaldi}
\affiliation{Surface Science Research
Centre, the University of Liverpool, Liverpoool L69 3BX, UK}
\author{Nicolas Lorente}
\affiliation{Institut de Ciencia de Materials Barcelona, ICMAB-CSIC,
Campus UAB, Bellaterra, Barcelona 08913, Spain}
\date{\today}%

\begin{abstract}
In low temperature scanning tunneling microscopy (STM) experiments a
cerium adatom on Ag(100) possesses two discrete states with
significantly different apparent heights. These atomic switches also
exhibit a Kondo-like feature in spectroscopy experiments. By
extensive theoretical simulations we find that this behavior is due
to diffusion of hydrogen from the surface onto the Ce adatom in
the presence of the STM tip field. The cerium adatom possesses vibrational
modes of very low energy (3--4meV) and very high efficiency ($\ge$
20\%), which are due to the large changes of Ce-states in the
presence of hydrogen. The atomic vibrations lead to a Kondo-like
feature at very low bias voltages. We predict that the same
low-frequency/high-efficiency modes can also be observed at
lanthanum adatoms.
\end{abstract}
%
%

\maketitle
Scanning tunneling microscopy and spectroscopy (STM/STS) are at
present the main tools to analyse the behavior of single atoms and
molecules at conducting surfaces. Experiments at very low
temperatures - typically around 4--5K - have contributed
considerably to our understanding of single atom contacts
\cite{berndt05}, Kondo-resonances and their signature in the
near-contact regime \cite{berndt07}, spin-flip excitations of single
atoms and atomic chains \cite{Heinrich04,Hirjibehedin06}, and
vibrational excitations of single molecules \cite{Stipe98}.
Recently, the changes of electronic properties or atomic
configurations due to field excitations were thoroughly investigated
in STM experiments \cite{dujardin06,stroscio06,harikumar06}. The
salient feature in these experiments is the ability to modify the
systems by varying the position and the field-intensity of the STM
tip. However, the electronic properties of atoms can also be
modified by the presence of hydrogen, as e.g. the
measurements of Gupta {\em et al.} showed for nominally clean
Cu(111) surfaces \cite{Gupta05}. They obtained spectroscopic data
with negative differential resistance, characteristic of vibrational
excitations of a molecule \cite{Gaudioso00}. Hydrogen itself, which
is quite ubiquitous in a low-temperature ultra-high vacuum (UHV)
environment, can only rarely be resolved in STM experiments
\cite{klein03}. Its effect on experimental scans has so far not been
studied in great detail.

The aim of this Report is to demonstrate the ability of manipulating
the position of atomic hydrogen at the nanometer scale by the field
of an STM tip and to determine its effect in spectroscopy
experiments at very low bias voltages. In this range, one typically
detects a Kondo resonance for magnetic adatoms
\cite{Fano61,berndt07}, which is due to the interaction of a
spin-state at a magnetic impurity with the conducting electrons of
the underlying metal. Kondo resonances have a very characteristic
signature, described by a Fano function
\cite{Madhavan01,Cornaglia03,Merino04}:
\begin{equation}
\frac{dI(V)}{dV} = A \frac{(\epsilon + q)^2}{1 + \epsilon^2} + B
\end{equation}
with $\epsilon = (eV - \epsilon_0)/\Gamma$. In this equation, $A$ is
the amplitude coefficient, $B$ is the background $dI/dV$ signal, $q$
is the Fano line shape parameter, $\epsilon_0$ is the energy shift
of the resonance from the Fermi level due to level repulsion between
the $d$-level and the Kondo resonance, and $\Gamma$ is the half
width of the resonance.
%
%
\begin{center}
\begin{figure}
\includegraphics[width=\columnwidth]{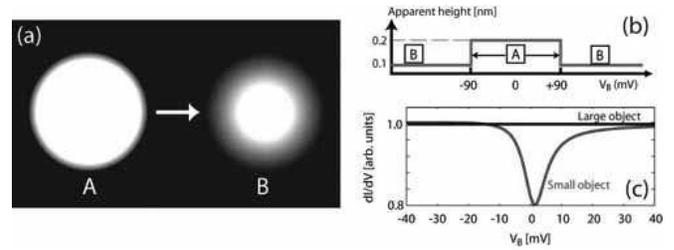}
  \caption{Experimental results from \cite{Ternes06}. (a) About 50\% of the Ce adatoms are seemingly
  unstable and show two different (large A and small B) appearence (tunneling parameters in the
  experiments: $V_T$=-90~mV, $I_T$=100~pA). (b) The unstable objects switch from
  a large protrusion of 200pm apparent height to a small protrusion of only
  100pm apparent height if the bias voltage exceeds 90mV in either polarization. (c) Correspondingly,
  a Kondo-like resonance is observed at the unstable adatoms, while the stable adatoms reveal a completely
  flat $dI/dV$ characteristic.}
\label{Ce-Ag(100)-exp}%
\end{figure}
\end{center}
In recent low-temperature experiments M. Ternes \cite{Ternes06}
found that adsorption of cerium atoms at Ag(100) surfaces leads to
two different atomic species at the silver surface in
low-temperature (4K) experiments. While most of the Ce adatoms
remained stable and showed an apparent height of about 200pm in all
STM scans, 5--50\% of the adatoms were highly unstable depending on
the applied bias voltage. For these objects it was found that a bias
of $\pm$90mV changes the apparent height from about 200pm to 100pm
(see Fig \ref{Ce-Ag(100)-exp}(a) and (b)). In this case the
structure remained stable as long as the STM tip remained at this
location, even if the bias voltage was subsequently reduced to zero
bias. The cause of this behavior could not be determined. In
addition, STS experiments at very low bias at the bistable objects
in its low-apparent-height state yielded a Fano function, which is
characteristic for Kondo resonances of magnetic impurities. The
$q$-value in the Fano fits was typically very low and in some cases
even negative (see Fig. \ref{Ce-Ag(100)-exp}(c)). However, the
change of conductance occurs at very low bias voltages, it is
therefore well below the range where inelastic effects due to atomic
vibrations are commonly observed.

To determine the cause of both effects, the change of apparent
height of the Ce adatom under certain tunneling conditions, and the
Fano-like behavior of the tunneling conductance at very low bias
voltages, we performed extensive electronic structure simulations
using density functional theory (DFT) and transport simulations for
elastic and inelastic tunneling processes. It will be seen that both
features have {\em one} common origin: the diffusion of hydrogen to the
and coadsorption at the
Ce adatom and the change of electronic properties as well as
vibrational efficiencies due to the coadsorption. Finally, we tested
the model for La adatoms, where a Kondo resonance is
certainly impossible. In this case we predict a similar behavior as
for Ce adatoms; a prediction which can be tested experimentally.
\begin{figure}[tb]
\begin{center}
\includegraphics[width=\columnwidth]{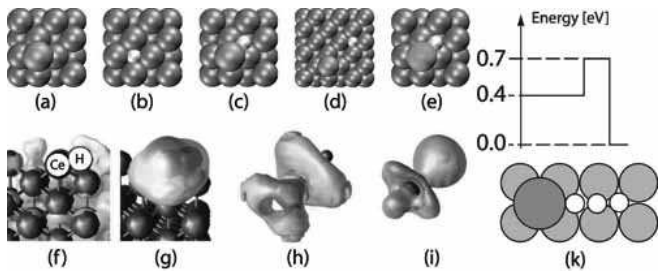}\hfill
\end{center}
 \caption{Configurations in the DFT simulations (a-e), charge transfer (f-i) and diffusion barriers (k).
 (a) Ce adatom on Ag(100). (b) H atom adsorbed close to the surface
 plane of Ag atoms. (c) Ce with coadsorbed H. (d) Ce with H at a neighboring lattice site. (e) La with coadsorbed H.
 Charge depletion (f) and accumulation (g) of Ag(100) due to coadsorption of Ce and
 H. The adsorption creates a polar charge distribution at the
 surface. (h) and (i) Same for Ce and H in the adsorbed geometry but
 without Ag. Here, the Ce atom looses charge to hydrogen. (k)
 Relative energy of coadsorbed hydrogen, hydrogen at a neighboring
 bridge site, and hydrogen in a fourfold hollow site of Ag(100).
} \label{Cells}
\end{figure}
%
%
%

In Figure \ref{Cells} (a) - (e) we show the groundstate
configurations of the calculated systems \cite{methods}. All single
adatoms adsorb at the fourfold hollow site. Coadsorbed hydrogen at
Ce and La is close to a bridge position of the silver surface. The
adsorption energy of hydrogen at Ag(100) is -2.84~eV (Configuration
(b)), coadsorbed at Ce it is smaller and only -2.43~eV
(Configuration (c)). From these results it can be concluded that the
groundstate of the system is given by Ce and H, adsorbed at fourfold
hollow sites. This result seems surprising at first view, because Ce
is routinely used in ultrahigh-vacuum chambers to remove hydrogen.
One would thus expect that the coadsorption of Ce and H should be
more favorable. To clarify this important point we also simulated
H$_2$, CeH, and CeH$_2$ in a vacuum, using the B3LYP functional.
Here, we obtained a binding energy of -4.10~eV for H$_2$, in good
agreement with experiments (-4.51~eV\cite{handbookCRC}), of -1.68~eV
for CeH, and of -3.50~eV for CeH$_2$. For CeH, the corresponding
value with standard functionals is -1.65~eV. From the difference in
the binding energies of CeH and H/CeAg(100) it is clear that a
significant proportion of the adsorption energy of coadsorbed
hydrogen is due to bonding to the Ag(100) surface and not to Ce.
This analysis leads to the unambiguous conclusion that coadsorbed
hydrogen at Ce/Ag(100) is {\em not} the groundstate of the system by
0.41~eV. Calculating the total energy of H and CeAg(100) for
hydrogen at an adjacent bridge site we find an energy barrier of
+0.29~eV. The relative energies for hydrogen diffusion from its
groundstate at the Ag(100) surface to Ce/Ag(100) are shown in Fig.
\ref{Cells}(k). Given the position of coadsorbed hydrogen and the
symmetry of the Ag(100) surface it is clear that four equivalent
pathways exist for the diffusion of hydrogen from its groundstate
position at the Ag(100) surface to Ce. However, it is not yet clear,
under which conditions hydrogen will coadsorb at Ce at all. There
have been a number of experiments, where diffusion of atomic
hydrogen due to the field of the STM tip has been
observed\cite{Stroscio91,klein03}. A straightforward calculation of
total energies under an applied dipole field could in principle
elucidate the change of adsorption energy due to an applied bias.
However, in case of CeH/Ag(100) DFT simulations did not arrive at a
converged solution, presumably due to the large size and complexity
of the system. To understand the effect of an external field on
CeH/Ag(100) we simulated the charge transfer by substracting the
charge density of the system components from the charge density of
the coupled system. We find that CeH leads to the occurrence of a
dipole moment in vertical direction since charge is removed from the
Ag(100) surface and accumulated at the position of CeH (Fig.
\ref{Cells}(f) and (g), the value of the contour is +0.001e/\AA$^3$,
and -0.006e/\AA$^3$, respectively). The change of total energy due
to a tip field ${\bf E}_{tip}$ is given by \cite{Jackson}:
%
%
\begin{equation}\label{dipole}
\Delta E = - {\bf P}_0 {\bf E}_{tip} - \alpha \chi_e |{\bf
E}_{tip}|^2
\end{equation}
Here, ${\bf P}_0$ is the residual dipole moment of CeH, $\alpha$ is
a constant, and $\chi_e$ is the electric susceptibility. Due to the
much higher number of valence electrons in Ce it can be expected
that $\chi_e$ is much larger for CeH/Ag(100) than for H/Ag(100).
From the charge difference contours we may conclude that the total
charge accumulated in the surface dipole is less than 0.05
electrons; the residual dipole, and consequently the asymmetric
changes of total energy with an applied tip field will be rather
small. The dominating term in the energy change should thus be due
to induced dipoles (the second term in Eq. \ref{dipole}), which is
symmetric with the applied field intensity. In this case we expect
that a certain threshold intensity in both polarizations will lower
the total energy of CeH/Ag(100) enough, so that it becomes more
favorable than the groundstate Ce+H/Ag(100). In this case hydrogen
will diffuse across the surface and coadsorb on Ce. Given that the
field for an atomically sharp tip is confined to a radius of less
than 1nm \cite{stokbro99}, diffusion will occur only in the
immediate vicinty of the STM tip and thus the position of the Ce
adatom. We may thus conclude that the higher susceptibility of Ce
compared to H leads to diffusion of hydrogen along the Ag(100)
surface. However, once CeH is formed, the diffusion barrier of
0.29eV will retain H at Ce, even if the bias voltage is decreased.
This conclusion is also in line with experimental data. From the
diffusion barrier of 0.29eV we may finally conclude, that a voltage
pulse of 0.3V should be sufficient for H to overcome the barrier and
diffuse along the surface. Given the high mobility of H even at low
temperature, we can thus predict that the observed object will
change again at higher positive bias voltages.
\begin{figure}[tb]
\begin{center}
\includegraphics[width=\columnwidth]{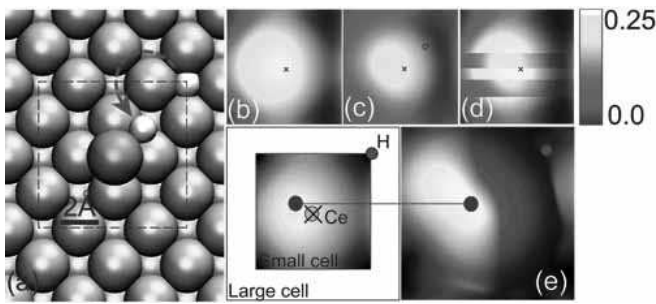}\hfill
\end{center}
 \caption{STM simulations of the switching feature observed on Ce adatoms. (a)
 at a  finite voltage increase the hydrogen atom diffuses across a barrier and attaches to the Ce adatom
 (see arrow). (b) Constant density contour for Ce adatom, and (c) for Ce adatom with coadsorbed
  hydrogen. (d) The contour changes its vertical distance by more than 1\AA. (e) Shift of the contour
  maximum due to hydrogen at a neigbouring lattice site. The maximum
shifts by about 1\AA \, compared to the contour at a clean Ag(100)
surface (yellow circle)
 if hydrogen is adsorbed at a neighboring lattice site (hydrogen: grey circle, new contour maximum:
 red circle).}
\label{sim_th}
\end{figure}
%
%
%

To determine the changes of the contours upon coadsorption of
hydrogen we performed STM simulations as described in the methods
section. The adatom appears as a bright protrusion with an apparent
height of about 2.5\AA(Fig. \ref{sim_th}(b)), and a diameter of
about 7\AA, in accordance with experimental images of the (i) large
stable, and (ii) large unstable atom. Subsequently, we simulated
coadsorbed hydrogen at the Ce adatom and determined the ensuing
constant density contours. Using the contour value for the bare
adatoms, we find that the density contours on top of Ce are now
1.2\AA \, closer to the surface, in line with experimental findings
(Fig. \ref{sim_th}(c) and (d))\cite{Ternes06}. The reason for the
lower contour is the decrease of density at the position of the Ce
adatom, as can be seen in the charge difference contours of CeH in
Fig. \ref{Cells}(h) and (i). From the simulations we may thus
conclude that the switching behavior of Ce adatoms is due to
reversible diffusion and coadsorption of hydrogen. We also simulated
the shift of a contour maximum, which should in principle be
detectable in STM scans, from an isolated Ce adatom to a system
where H is adsorbed at a neighboring hollow site (Fig.
\ref{Cells}(d)). Here, we find that the contour is displaced by
about 1\AA. Given the symmetry of the system, we expect four
equivalent shifts to occur in the experiments.
\begin{figure}[tb]
\begin{center}
\includegraphics[width=\columnwidth]{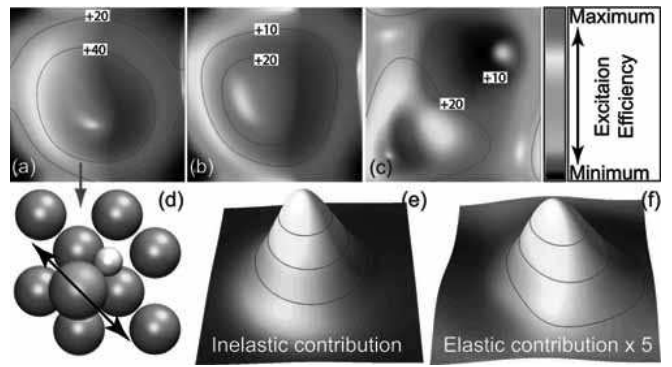}\hfill
\end{center}
\caption{Efficiency of two low-energy vibrational excitations for Ce
adatoms and coadsorbed hydrogen (a-b) and La with
 coadsorbed hydrogen (c). The colormap scale bar indicates the spatially
 resolved efficiency. The efficiency is positive in all cases,
 the mode with the highest efficiency, at 3.7~meV (frame (a)), is due to Ce-motion costrained
 by coadsorbed of hydrogen (see frame (d)). (e-f) High-efficiency mode of Ce/H.
 The inelastic efficiency is centered
 at the position of the Ce adatom (e)) and substantially larger than the elastic
 component (f).}
\label{eff_ce_h}
\end{figure}
%
%
%
An analysis of the $f$-states of Ce reveals that due to
their position at the Fermi level the Kondo temperature should be
well below the temperature in the experiments
(4K)\cite{berndt07}:  Kondo-resonances in this system are therefore not measurable.
This initial finding suggested to search for
a different origin. The only potential origin in this case are atomic vibrations.
In order to model low-energy vibrational excitations, we solved the
dynamical matrix for ionic motion of the Ce adatom, the hydrogen
adatom, and the silver surface layer keeping the other ionic
positions fixed. We find low-energy vibrational modes between 3--4~meV
 for both systems, Ce/Ag(100) and CeH/Ag(100). The reason for the
very low excitation energy is the very high mass of cerium. The
coupling to a vibrational mode of molecules depends on a
superposition of elastic and inelastic contributions to the
tunneling current, which yield a reduction and an increase in
current, respectively \cite{mii03,agrait02}. The total efficiency of
a vibrational mode thus depends significantly on the superposition
of these two contributions \cite{Lorente04}. Due to this effect, the
total efficiencies for the Ce/Ag(100) system in the low energy range
are close to zero for all calculated vibrational modes. However,
this changes drastically for systems with coadsorbed hydrogen. In
Fig. \ref{eff_ce_h}~(a-b) we show the total efficiencies  of two
low-energy vibrational modes as a function of position for
CeH/Ag(100). At the position of the Ce atom the efficiency in this
case is higher than 20\% (Fig. \ref{eff_ce_h}~(b)), or even 40\%
(Fig. \ref{eff_ce_h}~(a)). Given that the efficiency determined from
constant density contours is generally overestimated by a factor of
about two, the results for vibrational efficiencies under the
condition of hydrogen coadsorption agree very well with the
experiments (about 20\% in \cite{Ternes06}). At T=0~K the signature
of a vibration in the $dI/dV$ spectrum is a step in the conductance
at 3--4~mV in both polarizations; due to thermal broadening it will
be measured as an asymmetric feature: onset from 0~mV, reaching
its maximum of 20\% increase at 6--8~mV,
and nearly flat with a gradual decrease above 8~mV \cite{Gupta05}
(see Fig. \ref{Ce-Ag(100)-exp}(c)); a signature which is very
similar to a Kondo-resonance (dip at the Fermi level). The
coadsorption of hydrogen on La/Ag(100) leads to a vibrational
efficiency in the same range of 20--30\% at 3.5~meV
(Fig. \ref{eff_ce_h}~(c)). Pending the verification by STM experiments,
 we may conclude on the basis of simulations that
(i) the Kondo-like resonance is actually due to inelastic
excitations, and (ii) that these excitations will also be observed
for other adatoms in the presence of coadsorbed hydrogen. While for
bare Ce, the elastic components are nearly equal or even larger than
the inelastic ones, the coadsorption of hydrogen leads to a decrease
of the elastic efficiency, and, consequently to an increase of the
overall efficiency. An analysis of the high-efficiency phonon mode
at 3.7~meV reveals that it is due to a constrained vibration of
cerium. The mode is shown in Fig. \ref{eff_ce_h}~(d), the inelastic
and elastic efficiencies in Fig. \ref{eff_ce_h}~(e), and (f),
respectively. In the higher energy range above 5meV we found several
vibrational modes, both for Ce/Ag(100) and CeH/Ag(100). However, the
amplitude of the coupling for vibrational excitations is inversely
proportional to the excitation energy \cite{hellsing03}. While a low
frequency excitation with an efficiency of about 20\% can therefore
easily be detected at 3--4~meV, the ensuing increase in the $dI/dV$
spectrum will be less than 10\% for 8~meV, and below 5\% for 16~meV,
which corresponds roughly to the detection limit in the experiments.
Considering, in addition, that the superposition of elastic and
inelastic contributions in most cases makes these vibrations
invisible by scanning tunneling spectroscopy, it is reasonable to
assume that the higher energy modes will not leave a measurable
signature in the experimental scans.
%
%

To summarize, we found that dynamic processes of hydrogen adsorption
and diffusion, manipulated by the field of the STM tip, play a
pivotal role in low temperature experiments. These processes lead to
a bistable switching behavior of single adatoms. Here, vibrational
modes of  unusually low energies and unusually high efficiencies
have the same appearance as a typical Kondo-resonance. What seems
most interesting in this study is the possibility of engineering
subtle physical properties reversibly at the single-atom level
by field-induced diffusion processes.
%
%

{\bf Acknowledgements:} Helpful discussions with Wolf-Dieter
Schneider (EPFL Lausanne) and Markus Ternes (IBM Almaden) are
gratefully acknowledged. WAH is supported by the Royal Society, GT
is funded by EPSRC grant EP/C541898/1.
%
%

\bibliographystyle{apsrev}

\end{document}